\documentclass[a4paper,11pt]{article}

\pdfoutput=1
\usepackage[paper=letterpaper,margin=1.25in]{geometry}

\usepackage[hyperfootnotes=false, linktocpage=true, colorlinks, citecolor=blue, linkcolor=blue, urlcolor=blue, filecolor=blue]{hyperref}
\usepackage{graphicx}
\usepackage[subrefformat=parens,labelformat=parens]{subfig}
\usepackage{amsmath}
\usepackage{amssymb}
\usepackage{caption}
\usepackage{cite}
\usepackage{color}
\usepackage{enumitem}
\usepackage{xspace}
\usepackage{parskip}

\usepackage{amssymb,amsmath,epsfig}
\usepackage{bbm}
\usepackage{color}
\usepackage{cite}
\usepackage{epic}

\usepackage{multirow} % merger multi cells in a table
\numberwithin{equation}{section}
\newlength{\xtrawidth}
\newlength{\xtraheight}
\def\Z{\mathbb{Z}}
\def\R{\mathbb{R}}

\def\P{\mathbbm{P}\xspace}

\newcommand{\pt}{\partial}

\newcommand{\beq}{\begin{equation}}
\newcommand{\eeq}{\end{equation}}
\newcommand{\cL}{{\cal L}}

\newcommand{\cV}{{\cal V}}

\newcommand{\cO}{{\cal O}}
\newcommand{\cA}{{\cal A}}
\newcommand{\shortline}{\newline\vskip -7mm{\hbox to 2cm{\hrulefill}}\vskip 3mm}
\newcommand{\tX}{\ensuremath{\widetilde{X}}\xspace}
\newcommand{\tV}{\ensuremath{\widetilde{V}}\xspace}
\newcommand{\tW}{\ensuremath{\widetilde{W}}\xspace}

\newcommand{\qed}{\nobreak \ifvmode \relax \else
      \ifdim\lastskip<1.5em \hskip-\lastskip
      \hskip1.5em plus0em minus0.5em \fi \nobreak
      \vrule height0.75em width0.5em depth0.25em\fi}

\newcommand{\be}{\begin{equation}} % start the equation
\newcommand{\ee}{\end{equation}} % end the equation
\newcommand{\bi}{\begin{itemize}} % start the item env
\newcommand{\ei}{\end{itemize}} % end the item env 

\newcommand{\cpar}{\quad\quad} % indent of paragraph
\begin{document}
\begin{centering}
\vspace*{1.2cm}
{\LARGE \bf Heterotic Instanton Superpotentials from Complete\\[3mm] Intersection Calabi-Yau Manifolds}

\vspace{1cm}

{\bf Evgeny Buchbinder}${}^{1}$\footnote{evgeny.buchbinder@uwa.edu.au},
{\bf Andre Lukas}${}^{2}$\footnote{lukas@physics.ox.ac.uk},
{\bf Burt Ovrut}${}^{3}$\footnote{ovrut@elcapitan.hep.upenn.edu},
{\bf Fabian Ruehle}${}^{2}$\footnote{fabian.ruehle@physics.ox.ac.uk}

{\small
\vspace*{.5cm}
${}^{1}$Department of Physics, The University of Western Australia\\
35 Stirling Highway, Crawley WA 6009, Australia\\ [3mm]
${}^{2}$Rudolf Peierls Centre for Theoretical Physics, University of Oxford\\
  1 Keble Road, Oxford OX1 3NP, UK\\[0.3cm]
${}^{3}$Department of Physics and Astronomy, University of Pennsylvania\\ Philadelphia, PA 19104-6396, USA
}

\begin{abstract}\noindent
We study Pfaffians that appear in non-perturbative superpotential terms arising from worldsheet instantons in heterotic theories. A result by Beasley and Witten shows that these instanton contributions cancel among curves within a given homology class for Calabi-Yau manifolds that can be described as hypersurfaces or complete intersections in projective or toric ambient spaces. We provide a prescription that identifies all $\mathbb{P}^1$ curves in certain homology classes of complete intersection Calabi-Yau manifolds in products of projective spaces (CICYs) and cross-check our results by a comparison with the genus zero Gromov-Witten invariants. We then use this construction to study instanton superpotentials on those manifolds and their quotients. We identify a non-toric quotient of a non-favorable CICY with a single genus zero curve in a certain homology class, so that a cancellation \`a la Beasley-Witten is not possible. In another example, we study a non-toric quotient of a favorable CICY and check that the superpotential still vanishes. From this and related examples, we conjecture that the Beasley-Witten cancellation result can be extended to toric and non-toric quotients of CICYs, but can be avoided if the CICY is non-favorable.
\end{abstract}
\end{centering}

\newpage

\tableofcontents
\setcounter{footnote}{0}

%%%%%%%%%%%%%%%%%%%%%%%%%%%%%%%%%%%%%%%%%%%%%%%%%%%%%%%%%%%%%%%%%%%%%%%%%%%%%

\section{Introduction}
Compactifications of heterotic string theory have proven to provide ample realizations of models whose spectra and particle content closely resemble the Minimal Supersymmetric Standard Model (MSSM) or one of its extensions. In this paper, we focus on heterotic compactifications on smooth Calabi-Yau (CY) three-folds. An early example of a quasi-realistic model on a CY three-fold, based on a bundle with $SU(4)$ structure group in the observable sector \cite{Lukas:1997fg} of heterotic $M$-theory \cite{Lukas:1998yy}, can be found in Ref.~\cite{Braun:2005nv}. A large class of models can be constructed if the vector bundle is chosen to be a sum of line bundles \cite{Anderson:2011ns,Anderson:2012yf,Blaszczyk:2015zta,Nibbelink:2015vha}, since this considerably simplifies the otherwise hard task of checking supersymmetry of the bundle. All these models rely on dividing by a freely-acting discrete symmetry and CY three-folds with such symmetries indeed seem to be a necessary pre-requisite for realistic heterotic model building~\cite{Anderson:2014hia}. A large class of suitable examples, which we will focus on in the present paper, is provided by complete intersection CY manifolds (CICYs) in an ambient space which is a product of projective spaces. These CICYs have been classified in Ref.~\cite{Candelas:1987kf} and their freely-acting discrete symmetries have been identified in Ref.~\cite{Braun:2010vc}.

\cpar Apart from constructing models with an MSSM particle spectrum, moduli stabilization is another crucial step towards constructing realistic models. While a subset of the geometrical (that is, K\"ahler, complex structure and the dilaton) moduli  in heterotic CY models can be stabilised by flux~\cite{Anderson:2010mh,Anderson:2009nt}, this does not appear to be possible for all such moduli and, hence, non-perturbative effects such as worldsheet instantons and gaugino condensation are required. In fact, Ref.~\cite{Anderson:2011cza} presents a scenario which leads to the stabilisation of all geometrical moduli in certain heterotic CY models, based on both types of non-perturbative effects as well as flux. Be that as it may, these methods have not yet been applied to the stabilization of the non-geometric vector bundle moduli that arise in all realistic heterotic models. Although there is no general proof that the presence of instanton superpotentials is necessary for a successful stabilisation of moduli, they undoubtedly provide an important ingredient. Instanton superpotentials arise from strings wrapping genus zero curves in the CY manifold. However, even if such curves are present, a rather astonishing result of Beasley and Witten \cite{Beasley:2003fx} suggests that the instanton superpotential vanishes under fairly general assumptions, due to a cancellation of the contributions from curves within the same homology class. Understanding the scope and limitations of this result is clearly important for moduli stabilization and the construction of realistic models. The main purpose of the present paper is to study CICY manifolds from this point of view.

The proof of Beasley and Witten \cite{Beasley:2003fx} assumes that the CY three-fold satisfies certain geometric properties. Specifically, it assumes that the CY manifold is defined as a hypersurface or complete intersection in a projective or toric ambient space and that its K\"ahler class is ``favorable''. Favorable means that the K\"ahler class of the CICY descends from a K\"ahler class of the ambient space.
These conditions point to a number of ways in which the vanishing of the superpotential may be avoided:
\begin{enumerate}
  \item The ambient space is not toric.
  \item The CY manifold is not a complete intersection.
  \item The CY manifold is not favorable, that is, there are K\"ahler classes which do not descend from the ambient space.
\end{enumerate}   
All three possibilities can be realized within the context of CICY manifolds. Specifically, of the $7890$ CICY configuration matrices in the standard list of Ref.~\cite{Candelas:1987kf}, $2626$ are favourable while the others are not. While all CICY manifolds are defined as complete intersections in toric ambient spaces, the same is not necessarily true for their quotients by freely-acting discrete symmetries. We emphasise that realistic heterotic model building is based on these quotient manifolds which should, therefore, be the focus for discussing  the physical implications of instanton effects. It turns out that the list of freely-acting discrete symmetries for CICY manifolds in Ref.~\cite{Braun:2010vc} contains toric as well as non-toric symmetries. For the latter, the quotient of the ambient space is not toric nor does the quotient CY manifold have an obvious realization as a complete intersection. Consequently, CICY manifolds provide an interesting laboratory for studying instanton superpotentials.

Calculating the instanton superpotential contribution associated to a particular second homology class of the CY manifold requires knowledge of all the isolated, holomorphic genus zero curves in this class. The number of such curves can be determined from the Gromov-Witten invariants. However, finding the curves explicitly can be difficult and is one of the technical challenges in calculating instanton superpotentials. We will show how to find these curves for certain homology classes of CICY manifolds. Specifically, for a CICY defined in an ambient space ${\cal A}=\mathbb{P}^{n_1}\times \cdots\times\mathbb{P}^{n_m}$ we present a simple method, based on intersection theory, to determine genus zero curves in homology classes associated to $\mathbb{P}^1$ factors in the ambient space. By computing genus zero Gromov-Witten invariants we show, for the 7890 CICY manifolds in the standard list of Ref.~\cite{Candelas:1987kf} and all ambient space $\mathbb{P}^1$ factors, that this method provides all genus zero curves in those homology classes. This provides us with a large set of examples to explore heterotic instanton superpotentials.

While the list in Ref.~\cite{Candelas:1987kf} contains at least one realization for each topological type of CICY manifold, a given topological type often has many other realizations as a complete intersection in products of projective spaces. The full set of CICY configurations is, therefore, much larger than the standard list of Ref.~\cite{Candelas:1987kf}. We conjecture that our method extends to this full set of CICY configurations and provides the complete set of genus zero curves in all homology classes associated to ambient space $\mathbb{P}^1$ factors. 

Based on the above method to compute genus zero curves, we start exploring superpotentials on CICY manifolds by studying two specific examples. The first example is based on a non-favorable CICY manifold $\tilde{X}$ with a non-toric freely-acting $\mathbb{Z}_4$ symmetry and its quotient $X=\tilde{X}/\mathbb{Z}_4$. It turns out that the upstairs space $\tilde{X}$ has four genus zero curves in a specific homology class, associated to an ambient space $\mathbb{P}^1$ factor. Upon taking the $\mathbb{Z}_4$ quotient this descends to a single curve in the corresponding homology class of $X$. Hence, on the downstairs manifold $X$ there is no possibility of a cancellation between various genus zero curves and the superpotential must be non-vanishing. This example satisfies all three conditions above under which the Beasley-Witten result may be avoided. 

In order to gain a better intuition as to which of these three conditions is crucial, we study a second example. We consider a favorable CICY $\tilde{X}$ and its quotient $X=\tilde{X}/\mathbb{Z}_2$ by a non-toric $\mathbb{Z}_2$ symmetry. On $\tilde{X}$ we construct a vector bundle $\tilde{V}\rightarrow\tilde{X}$ as a double extension of line bundles which descends to a bundle $V\rightarrow{X}$. In a certain homology class of $\tilde{X}$, we find four genus zero curves which descend to two curves in the quotient manifold $X$. The upstairs superpotential contribution from this homology class is expected to vanish according to Beasley-Witten, which we verify explicitly. However, we also find that the downstairs superpotential contribution on $X$ vanishes due to a cancellation between the two curves. This result, which is confirmed by various similar examples, suggests that the crucial property required to avoid the Beasley-Witten vanishing result is non-favorability. 

The paper is organized as follows: In Section \ref{sec2} we introduce our notation, review the structure of non-perturbative superpotentials obtained from worldsheet instantons and the vanishing result of Beasley and Witten. In Section \ref{sec:InstantonNumbers} we explain the procedure to obtain genus zero curves as complete intersections and describe how to count these curves using intersection theory.  In Section \ref{sec:Examples} we present the two aforementioned examples. Conclusions and an outlook for future research directions follow in Section  \ref{sec:Conclusions}. In Appendix~\ref{sec:IsolatedCurves} we prove that the curves obtained from our method are indeed isolated.

%%%%%%%%%%%%%%%%%%%%%%%%%%%%%%%%%%%%%%%%%%%%%%%%%%%%%%%%%%%%%%%%%%%%%%%%%%%%%%

\section{Heterotic instanton superpotentials} \label{sec2}
Let us first introduce our notation. A general CY three-fold is denoted by $Y$ and a stable, holomorphic bundle on $Y$ by $U$. The symbol $C$ is used for any isolated, holomorphic genus zero curve, with $[C]$ its homology class. When we consider CICY manifolds, we write the projective ambient space as ${\cal A}=\mathbb{P}^{n_1}\times\cdots\times\mathbb{P}^{n_m}$, the (upstairs) CICY manifold as $\tilde{X}\subset{\cal A}$ and stable, holomorphic bundles on $\tilde{X}$ as $\tilde{V}$. If $\tilde{X}$ has a freely-acting discrete symmetry $\Gamma$ the quotient (downstairs) CY manifold is denoted by $X=\tilde{X}/\Gamma$  and stable, holomorphic bundles on $X$ as $V$.

%%%%%%%%%%%%%%%%%%%%%%%%%%%%%%%%%%%%%%%%%%%%%%%%%%%%%%%%%%%%%%%%%%%%%%%%%%%%%%%%

\subsection{The general structure}

%%%%%%%%%%%%%%%%%%%%%%%%%%%%%%%%%%%%%%%%%%%%%%%%%%%%%%%%%%%%%%%%%%%%%%%%%%%%%%%%%

We consider $E_8 \times E_8$ heterotic string theory or heterotic M-theory on a CY three-fold $Y$. As was extensively studied in a variety of 
papers~\cite{Dine:1986zy, Dine:1987bq, Becker:1995kb, Witten:1999eg, Harvey:1999as, Lima:2001jc},
the effective low-energy field theory contains a non-perturbative superpotential for moduli fields 
which is generated by worldsheet/open membrane instantons. The structure of the instantons as well as the structure of the ${\cal N}=1$ multiplets 
is slightly different in weakly and strongly coupled heterotic string theories but the superpotential has the same general form. For concreteness
we will discuss the weakly coupled case where the superpotential is generated by strings wrapping holomorphic isolated genus zero
curves $C$ in $Y$. The superpotential is then determined by the classical Euclidean worldsheet action evaluated on the instanton solution and by the one-loop determinants of the fluctuations around this solution. 
The general form of the superpotential induced by a string wrapping $C$ 
is~\cite{ Witten:1999eg}
\be 
W (C)= {\rm exp}\Big[ -\frac{A(C)}{2 \pi \alpha'} + i \int_C B \Big]
\frac{{\rm Pfaff}  ({\bar \pt}_{U_C (-1)})  }{ [ {\rm det}'  ({\bar \pt}_{{\cal O}_C} )]^2  [ {\rm det}  ({\bar \pt}_{{\cal O}_C(-1) } )]^2 }\,. 
\label{1.1}
\ee
The expression in the exponent is the classical Euclidean action evaluated on $C$. 
In the first term $A(C)$ is the area
of the curve given by 
\be 
A(C)= \int_C \omega_Y\,, 
\label{1.2}
\ee
where $\omega_Y$ is the K\"ahler form on $Y$. In the second term $B$ is the heterotic string $B$-field which in this expression can be taken 
to be a closed 2-form, $d B=0$. Let $\omega_I$ be a basis of $(1,1)$ forms on $Y$, where $I=1, \dots, h^{1,1} (Y)$, so that we can expand
\be 
\omega_Y=\sum_{I=1}^{h^{1,1}} t^I \omega_I\,, \qquad B=\sum_{I=1}^{h^{1,1}} \phi^I \omega_I\,. 
\label{1.3}
\ee
Defining the complexified K\"ahler moduli $T^I= \phi^I +i \frac{t^I}{2 \pi \alpha'}$, the exponential factor in Eq.~\eqref{1.1} becomes
\be 
e^{i \alpha_I(C) T^I}\,,  \qquad  \alpha_I(C) =\int_C \omega_I\,. 
\label{1.4}
\ee
The second factor in Eq.~\eqref{1.1} is the one-loop contribution which depends on the stable holomorphic vector bundle $U$ on $Y$.  The Pfaffian, ${\rm Pfaff}  ({\bar \pt}_{U_C (-1)})$, in the numerator is related to the Dirac operator on the curve $C$, twisted by the vector bundle $U_C (-1) = U|_{C} \otimes {\cal O}_{C} (-1)$.  It originates from integrating over the right-moving world-sheet fermions and is, in general, a homogeneous polynomial in the moduli of the vectors bundle $U$ and the complex structure moduli of $Y$. Explicit examples for the computation of Pfaffians can be found in Refs.~\cite{Buchbinder:2002ic, Buchbinder:2002pr, Buchbinder:2016rmw}. Finally,  ${\rm det}'  ({\bar \pt}_{{\cal O}_C} )$ and  ${\rm det}  ({\bar \pt}_{{\cal O}_C(-1) } )$ come from integrating over bosonic fluctuations. Since they are not important for the vanishing of the Pfaffians, we will not discuss them further; see Ref.~\cite{Witten:1999eg} for details.

In general, a given homology class of $Y$ contains more than one holomorphic isolated genus zero curve. The number of 
these curves is referred to as the (genus zero) Gromov-Witten invariant. All such curves in the same homology class have the same area, the same classical action and, hence, the same exponential factor in Eq.~\eqref{1.1}. However, the one-loop determinants are generally different. To find the superpotential contribution, $W([C])$, associated to the class $[C]$ we have to sum over all holomorphic, genus zero curves $C_j$ in this class. This leads to
\be 
W ([C])={\rm exp}\Big[ -\frac{A(C)}{2 \pi \alpha'} + i \int_C B \Big] \sum_{j=1}^{n_{[C]}}
\frac{{\rm Pfaff}  ({\bar \pt}_{U_{C_j} (-1)})  }{ [{\rm det}'  ({\bar \pt}_{{\cal O}_{C_j}} )]^2 [{\rm det}( {\bar \pt}_{{\cal O}_{C_j} (-1)})]^2}\,, 
\label{1.5}
\ee
where $n_{[C]}$ is the (genus zero) Gromov-Witten invariant of $[C]$. For the complete non-perturbative superpotential, $W$, we then have to sum over all homology classes; that is $W=\sum_{[C]}W([C])$.

%%%%%%%%%%%%%%%%%%%%%%%%%%%%%%%%%%%%%%%%%%%%%%%%%%%%%%%%%%%%%%%%%%%%%%%%%%%%%%

\subsection{The residue theorem of Beasley and Witten and its applicability}

%%%%%%%%%%%%%%%%%%%%%%%%%%%%%%%%%%%%%%%%%%%%%%%%%%%%%%%%%%%%%%%%%%%%%%
In Ref.~\cite{Beasley:2003fx} (also see the earlier papers~\cite{Distler:1986wm, Distler:1987ee, Silverstein:1995re, Basu:2003bq}) Beasley and Witten showed that under some rather general assumptions the sum~\eqref{1.5} vanishes. Let us briefly review their assumption. Let \tX be a complete intersection Calabi-Yau three-fold in the product of projective spaces\footnote{The results of Beasley and Witten are also expected to be valid for complete intersections in  toric spaces.} ${\cal A} = {\mathbb P}^{n_1} \times \dots \times  {\mathbb P}^{n_m}$. This means \tX is given by 
a set of  polynomial equations $p_1=p_2=\cdots = p_K=0$ with $\sum_{i=1}^m n_i - K =3$. Additionally, they assume that the vector bundle $\tV$ on $\tX$ is obtained as a restriction of a vector bundle ${\cV}$ on ${\cA}$, so that $\tV= {\cal V}|_{\tX}$. It was shown by Beasley and Witten that under these assumptions the sum~\eqref{1.5} vanishes for any homology class. 

As was pointed out in Ref.~\cite{Buchbinder:2016rmw}, the analysis of Beasley and Witten actually relies on the additional assumption that the K\"ahler form $\omega_{\tX}$ of $\tX$ is obtained as a restriction, $\omega_{\tX}=\omega_{{\cal A}}|_{\tX}$, of the ambient space K\"ahler form $\omega_{{\cal A}}$. If the CY manifold $\tX$ is favourable~\cite{Anderson:2008uw}, that is, if $h^{1, 1} ({\cal A})= h^{1, 1} (\tX)$ so that the entire second cohomology of $\tX$ descends from the ambient space, this assumption is indeed satisfied for all choices of K\"ahler form $\omega_{\tX}$. On the other hand, if $h^{1, 1} ({\cal A})<   h^{1, 1} (\tX)$ there may exist curves $C_j$ which have the same volumes as measured by restricted ambient space K\"ahler forms $\omega_{{\cal A}}|_{\tX}$ but different volumes as measured by K\"ahler forms $\omega_{\tX}$ which do not restrict from the ambient space. In this case, the statement of Beasley and Witten can still be applied~\cite{Buchbinder:2016rmw} to K\"ahler forms $\omega_{{\cal A}}|_{\tX}$ which descend from the ambient space and it implies the vanishing of the sum in Eq.~\eqref{1.5}. However, since the curves $C_j$ can have different volumes for choices of K\"ahler forms which do not descend from the ambient space, the exponential factors can be different and, hence, the superpotential does not vanish.

However, there is one more ingredient which was not considered in Ref.~\cite{Beasley:2003fx} and which can prevent the cancellation of 
individual instanton contributions in Eq.~\eqref{1.5}.  This ingredient is {\it discrete torsion}~\cite{Aspinwall:1994uj, Braun:2007tp, Buchbinder:2016rmw}.
In general, the second integer homology group of a CY manifold $Y$ is of the form 
\be 
H_2 (Y, {\Z})=  {\Z}^k \oplus G_{\rm tor}\,, \quad k >0\,, 
\label{1.9}
\ee
where $ {\Z}^k$ is the free part and $G_{\rm tor}$ is a discrete group which represents the torsion part. When  $G_{\rm tor}$ is non-trivial, curves with the same area (with respect to the a K\"ahler form $\omega_{Y}$) might be in different homology classes with respect to $G_{\rm tor}$ and, hence, be in different topological sectors. In the presence of the torsion, the expression~\eqref{1.5} is modified~\cite{Buchbinder:2016rmw} and becomes
\be 
W ([C])=e^{i \alpha_I (C) T^I} \sum_{j=1}^{n_{[C]}}
\frac{{\rm Pfaff}  ({\bar \pt}_{U_{C_j} (-1)})  }{ [{\rm det}'  ({\bar \pt}_{{\cal O}_{C_j}} )]^2
[ {\rm det} {\bar \pt}_{{\cal O}_{C_j} (-1)}]^2}\;  \chi  (C_j)\; ,
\label{1.10}
\ee
where $[C]$ is a homology class in $H_2 (Y, {\R})$ and the additional factor, $\chi  (C_j)$, is a character of $G_{\rm tor}$. Curves $C_j$ within the same torsion class come with the same factor $\chi(C_j)$. However, curves with different torsion classes may have different factors. While the sum in Eq.~\eqref{1.5} still cancels, the presence of the torsion factors means that the sum in Eq.~\eqref{1.10} can be non-vanishing.

%%%%%%%%%%%%%%%%%%%%%%%%%%%%%%%%%%%%%%%%%%%%%%%%%%%%%%%%%%%%%%%%%%%%%%%%

\section{Instanton numbers and genus zero curves} \label{sec:InstantonNumbers}
Instanton numbers in a given homology class of a Calabi-Yau manifold, that is Gromov-Witten invariants, can often be computed using known techniques~\cite{Candelas:1990rm,Hosono:1993qy}. However, for a calculation of the instanton superpotential, the $\P^1$ curves in the relevant homology class need to be known explicitly. Finding these curves is frequently not straightforward. In this section, we show that for certain homology classes of CICY three-folds, that is, the class of CY manifolds we are considering in this paper, there exists a systematic and simple procedure to find the $\P^1$ curves in certain homology classes explicitly. 

\subsection{Finding genus zero curves from complete intersections}
We recall that  a CICY three-fold is defined in an ambient space ${\cal A}=\P^{n_1}\times\cdots\times\P^{n_m}$ which consists of a product of projective spaces with dimensions $n_i$. It is given by the common zero locus of $K$ polynomials $p_a$, where $a=1,\ldots ,K$ and $K=\sum_{i=1}^m n_i-3$. 
The structure of the ambient space and the multi-degrees ${\bf q}_a=(q_a^1,\ldots ,q_a^m)^T$ of the polynomials $p_a$ are commonly encoded in the configuration matrix
\begin{equation}
 \tX\sim\left[\begin{array}{l|lll}\P^{n_1}&q_1^1&\cdots&q_K^1\\
                                            \vdots&\vdots&&\vdots\\
                                            \P^{n_m}&q_1^m&\cdots&q_K^m\end{array}\right]^{h^{1,1}(\tX),~h^{2,1}(\tX)}_{\eta(\tX)}\, .
\end{equation}                                            
The Hodge numbers of $\tX$ are usually attached as superscripts and the Euler number as a subscript. The Calabi-Yau condition is equivalent to $\sum_{a=1}^Kq_a^i=n_i+1$ for all rows $i=1,\ldots ,m$. We also introduce the standard K\"ahler forms $J_i$ on each $\P^{n_i}$, normalised such that $\int_{\P^{n_i}}J_i=1$. CICY three-folds have been classified in \cite{Candelas:1987kf} by finding a configuration matrix for each topological type and, in this way, a total of $7890$ configuration matrices have been identified. 

We would like to focus on cases where the ambient space contains at least one $\P^1$ factor (which is the case for $7762$ of the $7890$ configurations classified in \cite{Candelas:1987kf}) so that the ambient space has the form ${\cal A}=\P^1\times\tilde{\cal A}$, with $\tilde{\cal A}=\P^{n_2}\times\cdots\times\P^{n_m}$. In this case, the Calabi-Yau condition allows for two possible structures of the configuration matrix which (after a possible re-ordering of the defining polynomials) can be written as
\begin{equation}
\begin{array}{lllll}
 \mbox{type 1:}&&\tX_1&\sim&\left[\begin{array}{l|lllll}\P^1&1&1&0&\cdots&0\\
                                                  \tilde{\cal A}&\tilde{\bf q}_1&\tilde {\bf q}_2&\tilde{\bf q}_3&\cdots&\tilde{\bf q}_K\end{array}\right]\,,\\[6mm]
 \mbox{type 2:}&&\tX_2&\sim&\left[\begin{array}{l|llll}\P^1&2&0&\cdots&0\\
                                                  \tilde{\cal A}&\tilde{\bf q}_1&\tilde {\bf q}_2&\cdots&\tilde{\bf q}_K\end{array}\right]\,.
\end{array}\label{conf2}                                           
\end{equation}                                                  
Let us denote the homogeneous $\P^1$ coordinates by $[x_0:x_1]$ and the remaining coordinates of $\tilde{\cal A}$ by ${\bf y}$. For configuration matrices of type 1, the first two defining equations can be written as
\begin{equation}
\label{eq:Case1CICYEquation}
 p_1=x_0 \tilde{p}_1({\bf y})+x_1\hat{p}_1({\bf y})\, ,\qquad p_2=x_0 \tilde{p}_2({\bf y})+x_1\hat{p}_2({\bf y})\, , 
\end{equation}
where $\tilde{p}_i$ and $\hat{p}_i$, with $i=1,2$ are homogeneous polynomials of degree $\tilde{\bf q}_i$ in the coordinates ${\bf y}$ of $\tilde{\cal A}$. The remaining defining polynomials $p_i=p_i({\bf y})$ for $i=3,\ldots ,K$ are independent of the $\P^1$ coordinates. This means that the CICY $\tX_1$ contains a curve $\P^1\times{\bf y}$ for each point ${\bf y}\in\tilde{\cal A}$ which satisfies
\begin{equation}
 \tilde{p}_1({\bf y})= \hat{p}_1({\bf y})= \tilde{p}_2({\bf y})= \hat{p}_2({\bf y})=0\,,\quad p_i({\bf y})=0\,\mbox{ for all }\,i=3,\ldots ,K\, . \label{case1}
\end{equation} 
Note that these are $K+2$ equations on $\tilde{\cal A}$, a space of dimension $K+2$, so that the solution will generically be a finite number of points. 
Configuration matrices of type 2 in \eqref{conf2} can be discussed in a similar way. The first polynomial can now be written as
\begin{equation}
\label{eq:Case2CICYEquation}
p_1=x_0^2\tilde{p}_1({\bf y})+x_0x_1\hat{p}_1({\bf y})+x_1^2\bar{p}_1({\bf y})\, ,
\end{equation}
and the remaining polynomials, $p_i$ with $i=2,\ldots ,K$ only depend on the $\tilde{\cal A}$ coordinates ${\bf y}$. Hence, we have a curve $\P^1\times{\bf y}\in \tX_2$ for each point ${\bf y}\in\tilde{\cal A}$ which satisfies
\begin{equation}
 \tilde{p}_1({\bf y})= \hat{p}_1({\bf y})=\bar{p}_1({\bf y})=0\,,\quad p_i({\bf y})=0\mbox{ for all }i=2,\ldots ,K\, . \label{case2}
\end{equation}
As before, these are $K+2$ equations for the $K+2$ coordinates of $\tilde{\cal A}$ so generically the solution is a finite number of points.

To summarize, this means that we can obtain genus zero curves of the form $\mathbb{P}^1\times{\bf y}$  by finding the points $P_1=\{{\bf y}\}$ solving Eqs.~\eqref{case1} for type 1 cases and the points $P_2=\{{\bf y}\}$ solving~\eqref{case2} for type 2 cases. Note that, for a given choice of defining equations, this can be carried out explicitly. It may not be immediately obvious that these curves are isolated, but we have explicitly proven this in Appendix~\ref{sec:IsolatedCurves}. For a favorable CICY $\tilde{X}$, all curves obtained in this way are in the homology class dual to $J_1$. In the non-favorable case, the homology classes of all curves have a component dual to $J_1$ but they may differ by classes not obtained from the ambient space. (In particular, their volume is the same when measured by a K\"ahler form which descends from the ambient space.)

From Eqs.~\eqref{case1} and \eqref{case2} the point sets $P_1$ and $P_2$ can also be described by the configuration matrices
\begin{equation}
\begin{array}{lllll}
 \mbox{type 1:}&&P_1&\sim&\left[\begin{array}{l|lllllll}\tilde{\cal A}&\tilde{\bf q}_1&\tilde{\bf q}_1&\tilde{\bf q}_2&\tilde{\bf q}_2&\tilde{\bf q}_3&\cdots&\tilde{\bf q}_K\end{array}\right]\\[3mm]
 \mbox{type 2:}&&P_2&\sim& \left[\begin{array}{l|lllllll}\tilde{\cal A}&\tilde{\bf q}_1&\tilde{\bf q}_1&\tilde{\bf q}_1&\tilde{\bf q}_2&\cdots&\tilde{\bf q}_K\end{array}\right]
 \end{array}\, .\label{P12}
\end{equation}
The number of points in $P_1$ and $P_2$ can then be obtained by a standard intersection calculation based on these configuration matrices, that is, by carrying out the integrals
\begin{eqnarray}
|P_1|&=&\int_{\tilde{\cal A}}(\tilde{\bf q}_1\cdot{\bf J})^2\wedge (\tilde{\bf q}_2\cdot{\bf J})^2\bigwedge_{a=3}^K\tilde{\bf q}_a\cdot{\bf J}\\
|P_2|&=&\int_{\tilde{\cal A}}(\tilde{\bf q}_1\cdot{\bf J})^3\bigwedge_{a=2}^K\tilde{\bf q}_a\cdot{\bf J}\, ,
\end{eqnarray}
where ${\bf J}=(J_2,\ldots ,J_m)$ are the standard K\"ahler forms on $\tilde{\cal A}$. 

\subsection{Examples for calculating the Gromov-Witten invariants upstairs}
Let us carry out this analysis for two simple examples. Both examples are favorable so that all the curves we obtain lie in the same homology class in $\tX$. 

\subsubsection*{Example of type 1} 
We start with the CICY manifold (CICY 7858 in Ref.~\cite{Candelas:1987kf}) in the ambient space ${\cal A}=\P^1\times\P^4$ defined by
\begin{equation}
 \tX_1\sim\left[\begin{array}{l|ll}\P^1&1&1\\\P^4&3&2\end{array}\right]^{2,66}_{-128}\, .
\end{equation}
Clearly, this is a type 1 example and by comparing Eqs.~\eqref{conf2} and \eqref{P12} we see that the point set $P_1$ is described by the complete intersection
\begin{equation}
 P_1\sim\left[\begin{array}{l|llll}\P^4&3&3&2&2\end{array}\right]\; .
\end{equation}
With $J_2$ the standard K\"ahler form of $\P^4$, we find for the number of points
\begin{equation}
|P_1|=\int_{\P^4}(3J_2)^2\wedge (2J_2)^2=36\, .
\end{equation}
Hence, we find $36$ explicit curves $\P^1\times{\bf y}$ in the second homology class dual to $J_1$. 
A calculation of the Gromov-Witten invariant of this class, using the methods of Ref.~\cite{Hosono:1993qy}\footnote{A \texttt{Mathematica} implementation of their procedure is attached to their tex file on the arxiv.}, also leads to $36$. This shows that we have, in fact, 
found all the (genus zero) curves in this class via the method described above.

\subsubsection*{Example of type 2}
For a type 2 example, consider the well-known tetra-quadric (CICY 7862 in Ref.~\cite{Candelas:1987kf}) in the ambient space ${\cal A}=\P^1\times\P^1\times\P^1\times\P^1$; a favorable CICY defined by the configuration matrix
\begin{equation}
 \tX_2\sim\left[\begin{array}{l|l}\P^1&2\\\P^1&2\\\P^1&2\\\P^1&2\end{array}\right]^{4,68}_{-128}\, .
\end{equation} 
Comparing Eqs.~\eqref{conf2} and \eqref{P12}, the point set $P_2$ corresponds to the configuration
\begin{equation} 
 P_2\sim\left[\begin{array}{l|lll}\P^1&2&2&2\\\P^1&2&2&2\\\P^1&2&2&2\end{array}\right]\, ,
\end{equation}
and the number of points is given by
\begin{equation}
 |P_2|=\int_{(\P^1)^3}(2J_2+2J_3+2J_4)^3=48\, .
\end{equation}
This is the case since $J_i\wedge J_j\wedge J_k=2$ for $i,j,k$ mutually distinct and is zero otherwise. Therefore, the integral evaluates to $3 \cdot 2^3 \cdot 2=48$. Hence, we have found $48$ curves $\P^1\times{\bf y}$ in the second homology class of $\tilde{X}$ dual to $J_1$. Clearly, by symmetry, the other three $\P^1$ factors in the ambient space will lead to the same number of curves. A calculation of the Gromov-Witten invariant in those classes gives $48$. Hence, yet again, the method has produced all (genus zero) curves.\\[3mm]
We have performed the above intersection calculation for all $7890$ CICY manifolds in the list of Ref.~\cite{Candelas:1987kf} and all $\P^1$ factors in their ambient spaces. The resulting numbers of $\P^1$ curves in each homology class has been compared with the Gromov-Witten invariants obtained using~\cite{Hosono:1993qy}, and a perfect match has been found in all cases. Hence, at least for the $7890$ CICY manifolds in the standard list, all holomorphic isolated genus zero curves in homology classes associated to ambient space $\P^1$ factors can be computed explicitly. We expect that this remains true for all CICY three-fold configurations, including configuration matrices not contained in the standard list of Ref.~\cite{Candelas:1987kf} but equivalent to one of its entries. However, currently, we do not have a general proof. 

Given these explicit results for curves in certain homology classes of CICY manifolds, we have a large number of interesting and easily accessible examples for which to discuss the computation of heterotic instanton superpotentials. As it is, these Calabi-Yau manifolds are defined in ambient spaces which are products of projective spaces. Since they are known to not have discrete torsion, we conclude that for all favourable models the results of Ref.~\cite{Beasley:2003fx} apply and all $\P^1$ curves in a given homology class must sum to zero. Consequently, the instanton superpotential vanishes for these models. The situation might be different for non-favorable cases where the curves could lie in distinct classes with respect to the non-favorable part of $H_2(\tX,\Z)$.

\subsection{Freely acting symmetries and quotients}
\begin{figure}[t]
 \begin{tabular}{ccc}
  \subfloat[][Case \ref{enum:ActionType1}]{\includegraphics[height=.12\textwidth]{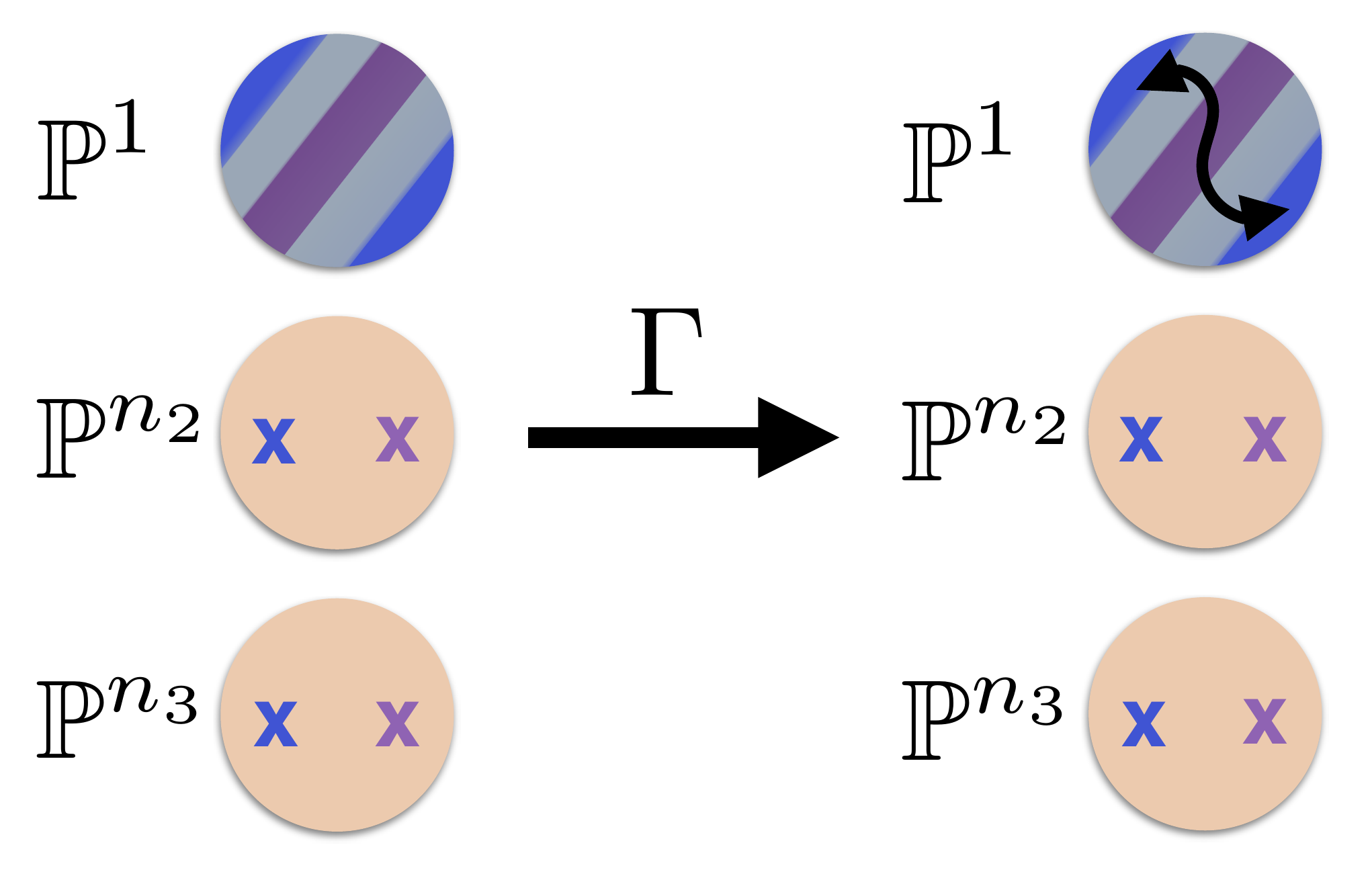}\label{fig:c1}}&\qquad\quad
  \subfloat[][Case \ref{enum:ActionType2}]{\includegraphics[height=.12\textwidth]{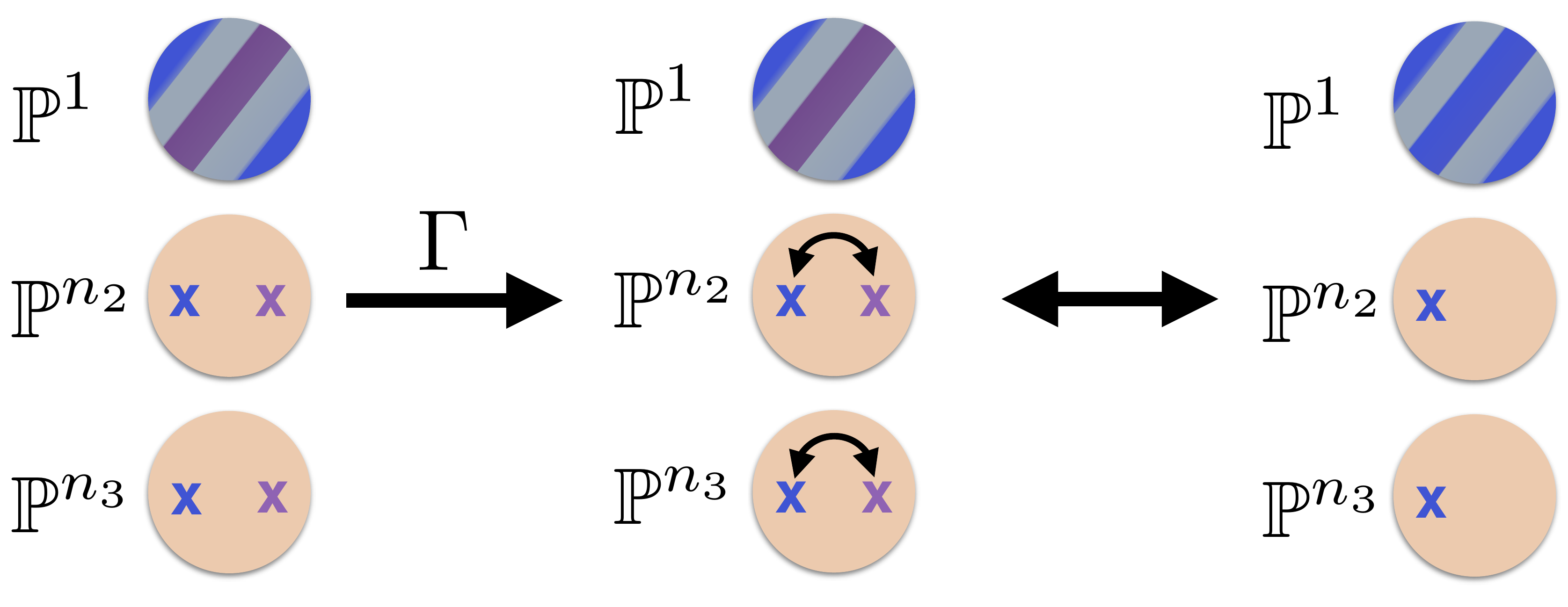}\label{fig:c2}}&\qquad\quad
  \subfloat[][Case \ref{enum:ActionType3}]{\includegraphics[height=.12\textwidth]{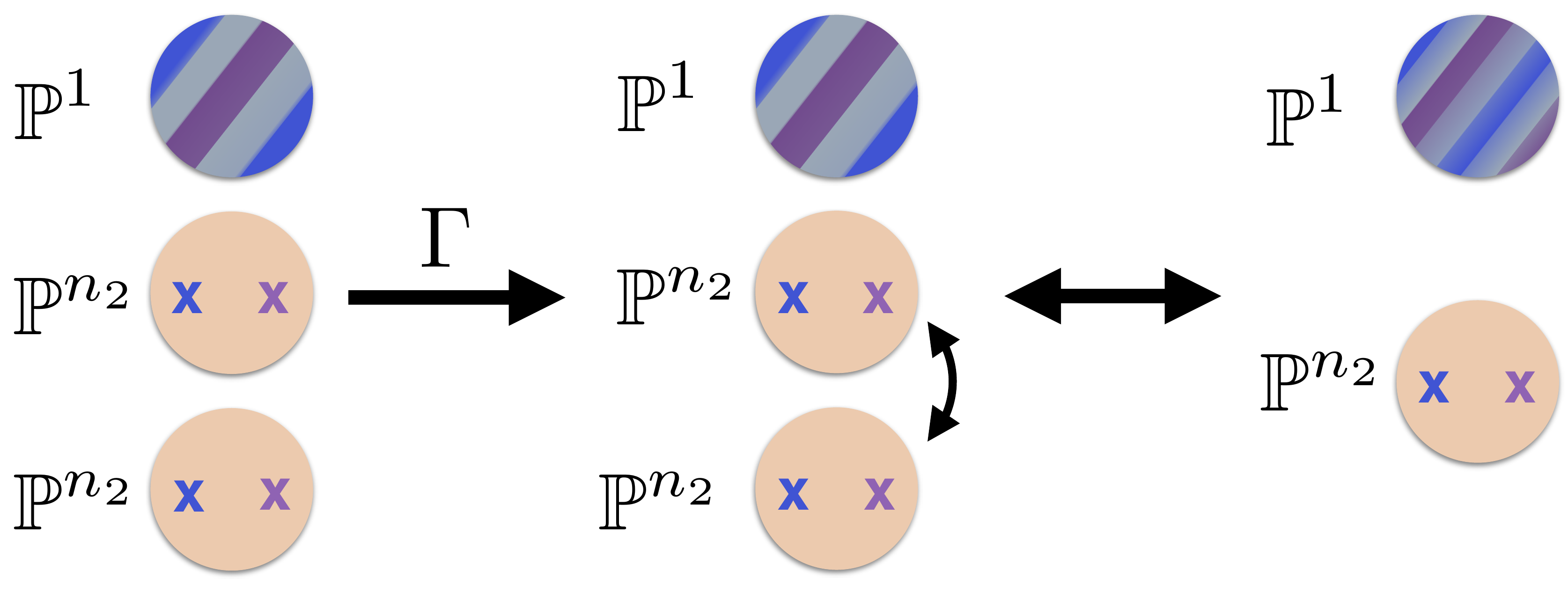}\label{fig:c3}}\\
  \multicolumn{3}{c}{\subfloat[][Case \ref{enum:ActionType4}]{\includegraphics[height=.12\textwidth]{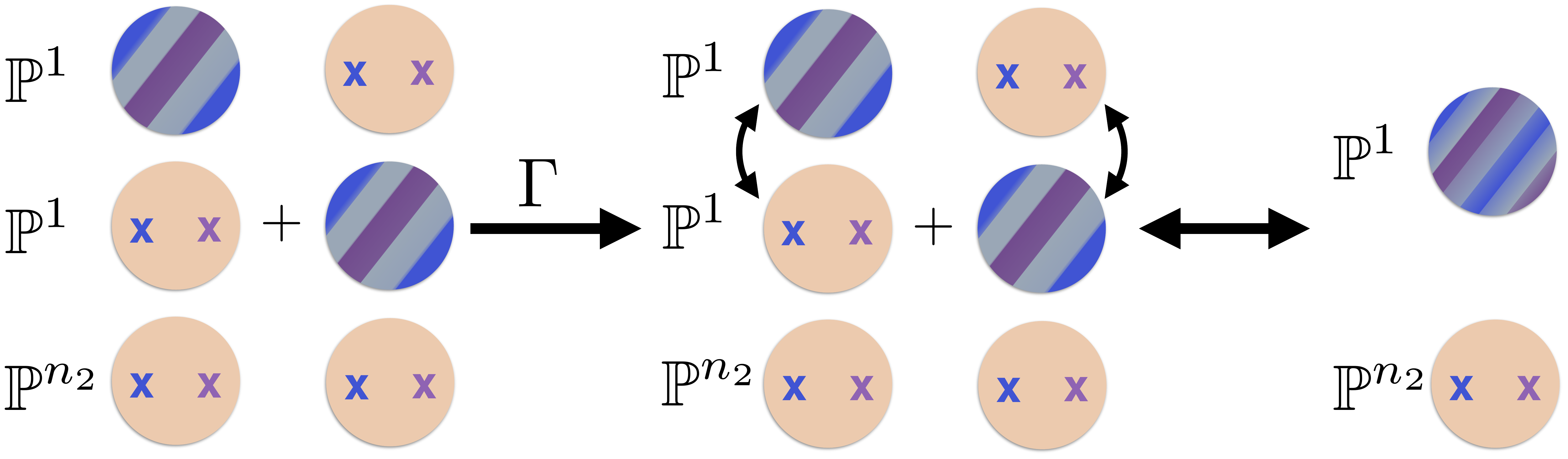}\label{fig:c4}}}
  \end{tabular}
  \caption{\sf Schematic depiction of the different types of free quotient actions \ref{enum:ActionType1} to \ref{enum:ActionType4}. Blue dots represent the $\mathbb{P}^1$ direction associated to the genus zero curves. The first two columns of each diagram indicate the action of the symmetry. The last column represents the resulting configuration in the quotient manifold.}
  \label{fig:FreeActionOnCurves}
\end{figure}
As explained in the introduction, we are not primarily interested in CICY manifolds $\tX$ themselves but, rather, in their quotients $X=\tX/\Gamma$ by freely-acting symmetries $\Gamma$. Such freely-acting symmetries of CICY manifolds have been classified in Ref.~\cite{Braun:2010vc}. If $\tX$ has a symmetry $\Gamma$, the genus zero curves identified above fall into orbits under the action of this symmetry. Upon taking the quotient, each orbit descends to a curve in $\tX/\Gamma$. The action of $\Gamma$ can be understood as a combination of the following four simple actions:
\begin{enumerate}[label=(\roman*)]
\item It acts on the homogeneous coordinates (by permuting them and/or multiplying them with phases) of the ambient space $\P^1$ factor associated to the genus zero curves Fig.~\ref{fig:FreeActionOnCurves}\subref{fig:c1}). \label{enum:ActionType1}
\item It acts on the homogeneous coordinates (by permuting them and/or multiplying them with phases) of the other ambient space factors $\P^{n_i}$, where $i>1$ (Fig.~\ref{fig:FreeActionOnCurves}\subref{fig:c2}). \label{enum:ActionType2}
\item It acts by permuting entire $\P^{n_i}$ factors, where $i>1$, that is, excluding the $\mathbb{P}^1$ factor associated to the genus zero curves (Fig.~\ref{fig:FreeActionOnCurves}\subref{fig:c3}). \label{enum:ActionType3}
\item It acts by permuting entire $\P^{n_i}$ factors, including the $\P^1$  factor associated to the genus zero curves (Fig.~\ref{fig:FreeActionOnCurves}\subref{fig:c4}). \label{enum:ActionType4}
\end{enumerate}
These basic actions are illustrated in Fig.~\ref{fig:FreeActionOnCurves}.

In case \ref{enum:ActionType1} only the parametrization of the genus zero curves is changed, but their counting is not affected at all. Indeed, Eqs.~\eqref{eq:Case1CICYEquation} and \eqref{eq:Case2CICYEquation} which count the number of genus zero curves are independent of the homogeneous $\P^1$ coordinates and are, hence, invariant under a symmetry acting in the $\P^1$ directions only.  In case \ref{enum:ActionType2} formerly independent solutions ${\bf y}$ to the equations \eqref{eq:Case1CICYEquation} or \eqref{eq:Case2CICYEquation} become identified under the action of the symmetry. This reduces the number of genus zero curves by the length of the orbit of the symmetry. Similar conclusions apply in case \ref{enum:ActionType3}. Finally, in case \ref{enum:ActionType4} the orbits of the symmetry consist of genus zero curves associated to different $\P^1$ ambient space factors.  It turns out that the length of the orbits always equals the order, $|\Gamma|$, of the discrete symmetry. 

The cases \ref{enum:ActionType3} and \ref{enum:ActionType4}, which permute entire ambient space factors always correspond to non-toric actions of $\Gamma$. Cases \ref{enum:ActionType1} and \ref{enum:ActionType2} can be toric or non-toric, depending on whether or not the action on the ambient space homogeneous coordinates can be diagonalized.  

If the genus zero curves associated to a certain $\mathbb{P}^1$ factor form a single orbit under the symmetry $\Gamma$, the quotient manifold only has a single curve in this homology class. In this case, there is only one contribution to the instanton superpotential from this downstairs homology class and a cancellation is impossible. We will present an explicit example in the next section which shows that this situation can indeed arise.

%%%%%%%%%%%%%%%%%%%%%%%%%%%%%%%%%%%%%%%%%%%%%%%%%%%%%%%%%%%%%%%%%%%%%%%%%%%%%%%

\section{Superpotential calculations}\label{sec:Examples}
In this section, we would like to present two explicit examples of superpotential calculations on CICY manifolds and their quotients. The first example involves a CICY manifold with freely-acting $\mathbb{Z}_4$ symmetry and four genus zero curves in a certain homology class which form a single orbit under the symmetry. Upon taking the quotient this results in a single genus zero curve and, hence, a non-vanishing superpotential contribution. The second example is for a CICY manifold with freely-acting $\mathbb{Z}_2$ symmetry, four genus zero curves in a certain upstairs class falling into two orbits and, hence, two resulting curves in the quotient. For a rank three bundle constructed by a double extension from line bundles, we show that the contributions from these two curves to the downstairs superpotential cancel. 

\subsection{Example 1: A quotient CY without Beasley-Witten cancellation}
\label{sec:ExampleBWAvoided}
%%%%%%%%%%%%%%%%%%%%%%%%%%%%%%%%%%%%%%%%%%%%%%%%%%%%%%%%%%%%%%%%%%%%%%%%%%%%%%

The example in question is for the compactification of the $E_8\times E_8$ (or $SO(32)$) heterotic string on a CICY manifold in the ambient 
space ${\cal A}=(\P^1)^3\times(\P^2)^2$ and specified by the configuration matrix
\begin{equation}
 \tX\sim\left[\begin{array}{l|llll}\P^1&1&1&0&0\\
                                           \P^1&0&0&0&2\\
                                           \P^1&0&0&2&0\\
                                           \P^2&1&0&0&2\\
                                           \P^2&0&1&2&0\end{array}\right]^{19,19}_0
\qquad
\begin{array}{l}x_0,x_1\\y_0,y_1\\z_0,z_1\\u_0,u_1,u_2\\v_0,v_1,v_2\end{array} \label{Xdef}                                           
\end{equation}     
This is CICY 30 in Ref.~\cite{Candelas:1987kf} and it is one of the possible realizations of the Schoen manifold. The notation for the homogeneous coordinates of each projective factor is indicated on the right-hand side of \eqref{Xdef}. 

For suitable choices of the defining polynomials, this CICY has a freely-acting $\Gamma=\mathbb{Z}_4$ symmetry. Its generator $\gamma$ acts linearly on the homogeneous coordinates $((x_0,x_1),(y_0,y_1),(z_0,z_1),(u_0,u_1,u_2),$ $(v_0,v_1,v_2))^T$ of the ambient space via the block matrix
\begin{equation}
 R(\gamma)=\left(\begin{array}{ccccc}i\sigma&0&0&0&0\\0&0&\sigma&0&0\\0&\mathbbm{1}_2&0&0&0\\0&0&0&0&s\\0&0&0&\mathbbm{1}_3&0\end{array}\right)\,,\quad \sigma={\rm diag}(1,-1)\,,\quad s={\rm diag}(1,-1,-1)\, . \label{Rg}
\end{equation}
This action is a combination of the action types \ref{enum:ActionType1} to \ref{enum:ActionType3}. Simultaneously, it acts on the defining equations $(p_1.\ldots ,p_4)^T$ via the matrix
\begin{equation}
 \rho(\gamma)=\left(
\begin{array}{cccc}
 0 & 1 & 0 & 0 \\
 1 & 0 & 0 & 0 \\
 0 & 0 & 0 & 1 \\
 0 & 0 & 1 & 0 \\
\end{array}
\right)\, .
\end{equation}
The resulting quotient $X=\tX/\Gamma$ has Hodge numbers $h^{1,1}(X)=h^{2,1}(X)=6$, see Ref.~\cite{Constantin:2016xlj}.

Let us now work out the number of holomorphic isolated genus zero curves associated to the first $\P^1$ factor in the configuration matrix \eqref{Xdef}. Clearly, this is a type 1 case and, from Eq.~\eqref{P12}, the complete intersection describing the points $P_1$ is given by
\begin{equation} 
 P_1\sim\left[\begin{array}{l|llllll}\P^1&0&0&0&0&0&2\\
                                                   \P^1&0&0&0&0&2&0\\
                                                   \P^2&1&1&0&0&0&2\\
                                                   \P^2&0&0&1&1&2&0\end{array}
                                                   \right]\, .
\end{equation}                                         
For the number of points we find
\begin{equation}
 |P_1|=\int_{(\P^1)^2\times (\P^2)^2}J_4^2\wedge J_5^2\wedge(2J_3+2J_5)\wedge(2J_2+2J_4)=4\, .
\end{equation} 
A similar calculation for the second and third $\P^1$ factor in \eqref{Xdef} leads to zero intersection points. As stated before, all these numbers match the 
Gromov-Witten invariants of the relevant homology classes. 

We would like to focus on the four curves associated to the first $\P^1$ factor. We denote these curves by 
$C_j=\P^1\times{\bf y}_j$, where $j=0,\ldots ,3$ and ${\bf y}_j$ are their locations in $\tilde{\cal A}=(\P^1)^2\times(\P^2)^2$. 
One can ask the question whether or not the contribution to the superpotential from these curves vanishes.
Since this example is {\it non-favorable} (note that 5=$h^{1,1 } ({\cal A}) < h^{1, 1} (\tX)=19$),
it is difficult to answer this question. Though these four curves have the same area with respect to a K\"ahler form obtained as restrictions from the ambient space, they might be in different homology classes in $\tX$. However, whether or not these four curves are in the same or in different 
homology classes in $\tX$ is not relevant for our purposes. In this paper, we are searching for other ways of ensuring that their contribution is non-vanishing. In fact, our primary interest is with the quotient $X$. 
\begin{figure}[t]
\centering
 \includegraphics[height=1cm]{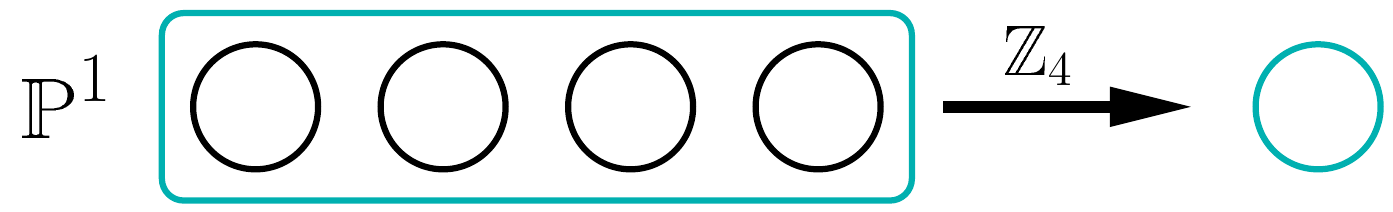}
 \caption{\sf This figure illustrates the identification of curves within the curve class corresponding to the first $\P^1$ ambient space factor. Upon modding out $\Gamma=\mathbbm{Z}_4$, the four different curves in this curve class are identified, leaving a curve class with Gromov-Witten invariant 1 on $X$.}
 \label{fig:DiscreteActionExample1}
\end{figure}

What is the situation for the quotient Calabi-Yau manifold $X=\tX/\Gamma$, where $\Gamma$ is the aforementioned freely-acting $\mathbb{Z}_4$ symmetry? The most general defining polynomials consistent with this symmetry are given by
\begin{align}
\begin{split}
 p_1&=x_0 \left(i c_4 u_1+i c_3 u_2\right)+x_1 \left(-i c_2 u_1-i c_1 u_2\right)\\
 p_2&=x_0 \left(c_4 v_1+c_3 v_2\right)+x_1 \left(c_2 v_1+c_1 v_2\right)\\
 p_3&=c_{14} v_0^2 z_0^2+c_{13} v_1^2 z_0^2+c_{11} v_2^2 z_0^2+c_{12} v_1 v_2
   z_0^2+c_{10} v_0 v_1 z_1 z_0+\\
   &\phantom{\;=\;}c_9 v_0 v_2 z_1 z_0+c_8 v_0^2 z_1^2+c_7 v_1^2 z_1^2+c_5 v_2^2 z_1^2+c_6 v_1 v_2 z_1^2\\
 p_4&=c_{14} u_0^2 y_0^2+c_{13} u_1^2 y_0^2+c_{11} u_2^2 y_0^2+c_{12} u_1 u_2
   y_0^2+c_{10} u_0 u_1 y_1 y_0+\\
   &\phantom{\;=\;}c_9 u_0 u_2 y_1 y_0+c_8 u_0^2 y_1^2+c_7 u_1^2 y_1^2+c_5 u_2^2 y_1^2+c_6 u_1 u_2 y_1^2 \, ,
\end{split}
\end{align}
where $c_1,\ldots ,c_{14}$ are arbitrary complex numbers which parametrize the choice of complex structure. Applying the general recipe~\eqref{case1} to these polynomials, we can find the four points ${\bf y}_j$ by solving
\begin{equation}
 \left(i c_4 u_1+i c_3 u_2\right)= \left(-i c_2 u_1-i c_1 u_2\right)=\left(c_4 v_1+c_3 v_2\right)=\left(c_2 v_1+c_1 v_2\right)=0
\end{equation}
along with $p_3=p_4=0$.  Explicitly, we find that
\begin{align}
\begin{split}
 {\bf y}_0&=(w_{2,0},-i w_{2,1},w_{2,0},-i w_{2,1},1,0,0,1,0,0)^T\\
 {\bf y}_1&=(w_{2,0},i w_{2,1},w_{2,0},-i w_{2,1},1,0,0,1,0,0)^T\\
 {\bf y}_2&=(w_{2,0},i w_{2,1},w_{2,0},i w_{2,1},1,0,0,1,0,0)^T\\
 {\bf y}_3&=(w_{2,0},-i w_{2,1},w_{2,0},i w_{2,1},1,0,0,1,0,0)^T\, ,
\end{split}
\end{align}
where $w_{2,0}=\sqrt{c_8}$ and $w_{2,1}=\sqrt{c_{14}}$. It is easy to verify, using the generator from Eq.~\eqref{Rg}, that ${\bf y}_j=R(\gamma)^{j}{\bf y}_0$ and, hence, that the four curves $C_j$ form one orbit under the action of the $\mathbb{Z}_4$ symmetry. As a result, these four curves are identified upon forming the quotient $X=\tX/\Gamma$. Hence, the corresponding downstairs homology class only contains a single holomorphic isolated genus zero curve $C$. Therefore, as long as a vector bundle $V$ on $X$ is chosen in such a way that ${\rm Pfaff}  ({\bar \pt}_{V_C (-1)}) $ is not identically zero, we a have a non-vanishing superpotential in this theory. 

For concreteness, we now present such a bundle. We start with an equivariant bundle $\tV$ on $\tX$ which then descends to a bundle $V$ on $X$. For simplicity, $\tV$ is chosen to be the sum of line bundles
\begin{equation}
  \tV=\cO_{\tX}(0,2,2,-1,-1)\oplus\cO_{\tX}(0,-2,-2,1,1)\; . \label{tVex}
\end{equation}  
It can be checked that this line bundle sum is equivariant under $\Gamma$ (note that each line bundle is clearly invariant) and that it satisfies the Bianchi Identities upon inclusion of NS5 branes. 
It is easy to check that it also allows for a solution to the slope zero conditions, if only the five favourable directions are taken into account. From a physical point of view, this might be problematic since a K\"ahler form $J$ induced from the ambient space might be on the boundary of the full K\"ahler cone. However, using the fact that the Schoen manifold can be written as a blowup of a $T^6/(\Z_2\times\Z_2)$ orbifold \cite{Nibbelink:2012de}, we can use the techniques of Ref.~\cite{Blaszczyk:2011hs,Nibbelink:2016wms} to match the CICY description to the resolved orbifold description. In the latter, we have an explicit realization of all 19 divisor classes. Using this map, we can check that the Hermitian Yang-Mills equations have a solution inside the full K\"ahler cone of the Schoen manifold. In summary, the line bundle sum $\tV$ does indeed provide a consistent choice and it descends to a line bundle sum $V$ on the quotient manifold $X$. 

Recall that the genus zero curves under consideration are associated to the first $\mathbb{P}^1$ factor. Since the corresponding first entries in the line bundle sum~\eqref{tVex} are zero, it follows that
\begin{equation}
 V_C(-1)=V|_C\otimes{\cal O}_{\mathbb{P}^1}(-1)={\cal O}_{\mathbb{P}^1}(-1)^{\oplus 2}\; .
\end{equation} 
Since ${\cal O}_{\mathbb{P}^1}(-1)$ does not have sections, the resulting Pfaffian ${\rm Pfaff}  ({\bar \pt}_{V_C (-1)}) $ is not identically zero. Note that, in this case, the non-vanishing of the superpotential does not rely on subtle geometric features such as torsion in $H_2(X, {\mathbb Z})$.

How does this example avoid the no-go theorem of Ref.~\cite{Beasley:2003fx}? First note that the downstairs ambient space ${\cal A}/\mathbb{Z}_4$ is 
singular (but the singularities do not, generically, intersect the Calabi-Yau manifold $X$). Furthermore, the $\mathbb{Z}_4$ symmetry does not act in a toric way on ${\cal A}$, as the generator~\eqref{Rg} shows. Hence, the downstairs ambient space (after blowing up the singularities) is neither a product of projective spaces nor does it have a toric description. Further, while the upstairs space \tX is a CICY manifold the same is not the case for the quotient CY $X$. Finally, neither $\tX$ nor $X$ are favourable.  However, all these assumptions enter in the proof of the vanishing statement in Ref.~\cite{Beasley:2003fx}. 

To develop a better intuition for which of these assumptions are crucial for the vanishing of the superpotential, we study a second example which also involves a quotient of a CICY manifold by a non-toric symmetry, but with both $\tX$ and $X$ favorable manifolds.

\subsection{Example 2: A quotient CY with Beasley-Witten cancellation}
The analysis of this example will follow the ideas of Ref.~\cite{Buchbinder:2016rmw}. It is based on the CICY manifold 6804 with configuration matrix
\begin{align}
 \tX\sim\left[
  \begin{array}{lllll}
 	\P^1		&1&1&0&0\\
    \P^1		&0&0&2&0\\
    \P^1		&0&0&0&2\\
    \P^2		&1&0&1&1\\
    \P^2		&0&1&1&1\\ 
   \end{array}\right]^{5,37}_{-64}
\qquad
\begin{array}{l}x_0,x_1\\y_0,y_1\\z_0,z_1\\u_0,u_1,u_2\\v_0,v_1,v_2\end{array} \,.
\label{eq:ConfCICY6804}                            
\end{align}
The ambient space K\"ahler forms (as well as their restrictions to the CY) are denoted by $J_i$, where $i=1,\ldots , 5$ and the second Chern class of the tangent bundle, relative to a basis dual to $J_i$, is given by 
\begin{equation}
 c_{2,i}(T\tX)=(24,24,24,36,36)\; . \label{c2TX}
\end{equation} 
The non-vanishing triple intersection numbers $\kappa_{ijk}=\int_{\tX}J_i\wedge J_j\wedge J_k$ are explicitly given by
\begin{align}
\begin{split}
\kappa_{1,2,3}&=\kappa_{1,2,4}=\kappa_{1,2,5}=\kappa_{1,3,4}=\kappa_{1,3,5}=2\,,\qquad \kappa_{1,4,5}=\kappa_{2,4,5}=\kappa_{3,4,5}=4\,,\\
\kappa_{2,3,4}&=\kappa_{2,3,5}=3\,,\qquad \kappa_{2,2,4}=\kappa_{2,2,5}=\kappa_{3,4,4}=\kappa_{3,5,5}=2\,,\qquad \kappa_{4,4,5}=\kappa_{4,5,5}=2\; . 
\end{split}\label{isec}
\end{align}
This manifold has a freely-acting $\Gamma=\mathbb{Z}_2$ symmetry with generator $\gamma$. Its action on the homogeneous ambient space coordinates is given by
\begin{align}
R(\gamma)=
\left(
\begin{array}{cc|cc|cc|ccc|ccc}
 -1 & 0 & 0 & 0 & 0 & 0 & 0 & 0 & 0 & 0 & 0 & 0 \\
 0 & 1 & 0 & 0 & 0 & 0 & 0 & 0 & 0 & 0 & 0 & 0 \\
 \hline
 0 & 0 & -1 & 0 & 0 & 0 & 0 & 0 & 0 & 0 & 0 & 0 \\
 0 & 0 & 0 & 1 & 0 & 0 & 0 & 0 & 0 & 0 & 0 & 0 \\
 \hline
 0 & 0 & 0 & 0 & -1 & 0 & 0 & 0 & 0 & 0 & 0 & 0 \\
 0 & 0 & 0 & 0 & 0 & 1 & 0 & 0 & 0 & 0 & 0 & 0 \\
 \hline
 0 & 0 & 0 & 0 & 0 & 0 & 0 & 0 & 0 & 1 & 0 & 0 \\
 0 & 0 & 0 & 0 & 0 & 0 & 0 & 0 & 0 & 0 & 1 & 0 \\
 0 & 0 & 0 & 0 & 0 & 0 & 0 & 0 & 0 & 0 & 0 & 1 \\
 \hline
 0 & 0 & 0 & 0 & 0 & 0 & 1 & 0 & 0 & 0 & 0 & 0 \\
 0 & 0 & 0 & 0 & 0 & 0 & 0 & 1 & 0 & 0 & 0 & 0 \\
 0 & 0 & 0 & 0 & 0 & 0 & 0 & 0 & 1 & 0 & 0 & 0 \\
\end{array}
\right)\,.\label{Z2gen}
\end{align}
This amounts to the same toric action on all three ambient space $\mathbb{P}^1$ factors and a simultaneous swap of the two $\mathbb{P}^2$ factors. Altogether, the action of this symmetry is evidently non-toric.

The genus zero Gromov-Witten invariants in this case are
\begin{align}
n_{(1,0,0,0,0)}=4\,,\quad n_{(0,1,0,0,0)}=12\,,\quad n_{(0,0,1,0,0)}=12\,,\quad n_{(0,0,0,1,0)}=32\,,\quad n_{(0,0,0,0,1)}=32\,.
\end{align}
For the purposes of this discussion, we will focus on the first $\mathbb{P}^1$ ambient space factor with four genus zero curves.

For the vector bundle $\tilde{V}$, we would like to consider a bundle with non-Abelian structure group so that the Pfaffian is a non-trivial function of the bundle moduli. Given that we are not computing the overall factor of the one-loop contribution in Eq.~\eqref{1.5}, such a non-trivial bundle moduli dependence is essential in order to check for the cancellation between contributions. Specifically, we define the rank three bundle $\tilde{V}$ as a double extension
\begin{align}
\label{eq:ExtensionSES}
\begin{array}{c@{\;}c@{\;}c@{\;}c@{\;}c@{\;}}
0&\longrightarrow L_1	&\longrightarrow \tW &\longrightarrow L_2&\longrightarrow 0\\
0&\longrightarrow \tW		&\longrightarrow \tV  &\longrightarrow L_3&\longrightarrow 0\; ,
\end{array}
\end{align} 
where $\tW$ is a rank two  auxiliary bundle and the line bundles $L_i$ are defined as restrictions, $L_i={\cal L}_i|_{\tX}$ of ambient space line bundles ${\cal L}_i$. Furthermore, these line bundle are chosen to satisfy $L_1\otimes L_2\otimes L_3=\cO_{\tX}$ to ensure that $\tV$ defines an SU(3) rather than a U(3) bundle. Explicitly, they are chosen as
\begin{align}
L_1=\cO_{\tX}(-2, 1, 1, 0, 0)\,,\qquad L_2=\cO_{\tX}(0, 1, -2, 0, 0)\,,\qquad L_3=\cO_{\tX}(2, -2, 1, 0, 0)\,.
\end{align}
Since we want to mod out a freely acting symmetry $\Gamma=\mathbbm{Z}_2$ later, we first check that the bundle is equivariant with respect to this symmetry. This is ensured since all line bundles $\cL_i$, $i=1,2,3$ are equivariant (and provided that we choose suitable extension classes). Note that $\Gamma$ will act by swapping the two $\P^2$ coordinates, which is why we have chosen the first Chern class in those directions to be the same.

Does this choice of line bundles lead to a non-trivial moduli space of extension bundles? To answer this question, we have to compute $H^1(\tX,\tW\otimes L_3^*)$; that is, the space of extension bundles $\tV$. To do this, we twist the first short exact sequence in \eqref{eq:ExtensionSES} by $L_3^*$ and consider the associated induced long exact sequence in cohomology,
\begin{align}
\label{eq:ExtensionLES}
\begin{array}{l@{\;}l@{\;}l@{\;}l@{\;}}
0&\longrightarrow H^0(\tX,L_1\otimes L_3^*)	&\longrightarrow H^0(\tX,\tW\otimes L_3^*) &\longrightarrow H^0(\tX,L_2\otimes L_3^*)	\\
  &\longrightarrow H^1(\tX,L_1\otimes L_3^*)	&\longrightarrow \underline{H^1(\tX,\tW\otimes L_3^*)} &\longrightarrow H^1(\tX,L_2\otimes L_3^*)	\\
  &\longrightarrow H^2(\tX,L_1\otimes L_3^*)	&\longrightarrow \ldots   &	
\end{array}\; .
\end{align} 
We are interested in the underlined term in this sequence. Given the cohomologies
\begin{equation}
h^\bullet(\tX, L_1\otimes L_3^*)=(0,12,10,0)\;,\qquad h^\bullet(\tX,L_2\otimes L_3^*)=(0,0,32,0)\;, \label{Licoh}
\end{equation}
computed using the methods described in Ref.~\cite{CICYPackage}, we conclude that
\begin{equation}
  h^1(\tX,\tW\otimes L_3^*)=12\; . \label{extdim}
\end{equation}
Hence, the extension space of bundles $\tV$ is indeed non-trivial and $12$-dimensional. For later purposes, it is also useful to note, using the Koszul sequence, that
\begin{equation}
  H^1(\tX,\tW\otimes\cL_3^*)\cong H^1(\tX, L_1\otimes L_3^*)\cong H^1({\cal A},{\cal L}_1\otimes{\cal L}_3^*)=H^1({\cal A},{\cal O}_{\cal A}(-4,3,0,0,0))\; .
\end{equation}  
The last expression provides us with an explicit way of writing down an arbitrary extension class in terms of ambient space coordinates. After Serre dualizing in the direction of the first $\mathbb{P}^1$ factor, such an arbitrary extension class can be written as a polynomials with multi-degree $(2,3,0,0,0)$; that is, as 
\begin{align}
\label{eq:BasisH1}
v=x_0^2 f_1(\mathbf{y})+x_0 x_1 f_2(\mathbf{y})+x_1^2 f_3(\mathbf{y})\; ,
\end{align}
with cubics $f_i(\mathbf{y})$ in the coordinates of the second $\mathbb{P}^1$ factor.  These cubics can be written explicitly as
\begin{align}
\begin{split}
 f_1(\mathbf{y})&=a_0 y_0^3 + a_1 y_0^2 y_1 + a_2 y_0 y_1^2 + a_3 y_1^3\,,\\
 f_2(\mathbf{y})&=b_0 y_0^3 + b_1 y_0^2 y_1 + b_2 y_0 y_1^2 + b_3 y_1^3\,,\\
 f_3(\mathbf{y})&=c_0 y_0^3 + c_1 y_0^2 y_1 + c_2 y_0 y_1^2 + c_3 y_1^3\,,
 \end{split}\label{fidef}
\end{align}
where  $a_k,b_k,c_k$, with $k=0,\ldots,3$, are coefficients. Note that the total number of these coefficients is $12$, in accordance with Eq.~\eqref{extdim}.

Next, we check that the SU(3) bundle $\tV$ can satisfy the Bianchi identities with NS5 branes; that is,  $c_2(T\tX)-c_2(\tV)=[\widetilde{M}]$ for some effective curve class $[\widetilde{M}]$. From 
\begin{equation}
 c_2(\tV)=\frac{1}{2}\sum_{i=1}^3c_1(L_i)^2
\end{equation}
and the intersection numbers~\eqref{isec}, we find that
\begin{align}
c_{2,i}(\tV)=(6,0,12,21,21)\; 
\end{align}
relative to the basis dual to $J_i$. Comparison with the second Chern class~\eqref{c2TX} of the tangent bundle shows that the Bianchi identity can indeed be satisfied by wrapping five branes on a curve $\tilde{M}$ with class $[\tilde{M}]_i=(18,24,12,15,15)$. 
 
To show that the bundle $\tV$ is poly-stable, we begin at the split locus of the extensions where $\tV \cong L_1\oplus L_2\oplus L_3$. It is easy to verify, using the intersection numbers~\eqref{isec}, that the slopes $\mu(L_i)=\int_{\tX}J\wedge J\wedge c_1(L_i)$ of the three line bundles vanish simultaneously at a locus in K\"ahler moduli space. Hence, on this locus, $\tV$ is poly-stable. In order to show that $\tV$ is poly-stable away from the split locus, we have to show that all rank one and two sub-sheaves $\widetilde{S}$ injecting into $\tV$ have a slope $\mu(\widetilde{S})$ satisfying $\mu(\widetilde{S})<\mu(\tV)=0$. This is rather tedious but can indeed be checked explicitly. Alternatively, we note that the cohomologies \eqref{Licoh} imply the existence of sufficiently general matter field terms in the low-energy D-terms, so that supersymmetric, D-flat directions away from the split locus clearly exist. 

With this we can finally work out the Pfaffian. As explained in Refs.~\cite{Witten:1996bn,Buchbinder:2002ic,Buchbinder:2002pr}, the Pfaffian on a holomorphic isolated genus zero curve $C$ vanishes if and only if $h^0(C, \tV|_C\otimes\cO_C(-1))\neq0$. While the dimension of this cohomology is zero generically, it can jump on a special locus in bundle moduli space. The Pfaffian for a curve $C_j=\mathbb{P}\times {\bf y}_j$ with homology class $[C]$ is proportional to the equation describing this special locus; typically a determinant of a certain matrix. For the case at hand, this matrix is given by~\cite{Buchbinder:2016rmw} 
\begin{align}
\label{eq:PfaffianDet}
d_j=\text{det}\left[\begin{pmatrix}
 f_1(\mathbf{y}_j) & f_2(\mathbf{y}_j)\\
 f_2(\mathbf{y}_j) & f_3(\mathbf{y}_j)
\end{pmatrix}
\right]\,,
\end{align}
where the polynomials $f_i$ have been defined in Eq.~\eqref{fidef}. Recall that we are focusing on the four genus zero curves associated to the first $\mathbb{P}^1$ ambient space factor and that the points ${\bf y}_i$ are their locations in the remaining ambient space factors $\tilde{A}=(\mathbb{P}^1)^2\times (\mathbb{P}^2)^2$. These points can be explicitly determined following the procedure described in Section \ref{sec:InstantonNumbers}. Anticipating the $\Gamma=\mathbb{Z}_2$ to be chosen later on, we will do this for the most general set of $\mathbb{Z}_2$ invariant defining equations for $\tX$. This leads to the four points
\begin{eqnarray}
 {\bf y}_0&=&(w_{2,0},1,w_{3,0},1,w_{4,0},w_{4,1},1,w_{4,0},w_{4,1},1)^T \nonumber \\
 {\bf y}_1&=&(w_{2,0},1,-w_{3,0},1,w_{4,0},w_{4,1},1,w_{4,0},w_{4,1},1)^T \nonumber\\
 {\bf y}_2&=&(-w_{2,0},1,w_{3,0},1,w_{4,0},w_{4,1},1,w_{4,0},w_{4,1},1)^T \nonumber \\
 {\bf y}_3&=&(-w_{2,0},1,-w_{3,0},1,w_{4,0},w_{4,1},1,w_{4,0},w_{4,1},1)^T \, , 
\label{eq:P1PointsCICY6804}
\end{eqnarray}
where the $w_{a,b}$ are known, but complicated, functions of the complex structure moduli appearing in the defining equations.

The Beasley-Witten vanishing theorem now tells us that all four contributions in \eqref{eq:PfaffianDet}, with the  above points ${\bf y}_i$ inserted, sum to zero. As remarked previously, the present method does not compute the relative factors between the summands. So all one can do is to check whether the four polynomials $d_j$ are linearly dependent, that is, whether there are $k_j\in\mathbbm{C}$, independent of the bundle moduli $a_k,b_k,c_k$, such that
\begin{align}
\sum_{j=0}^{3}k_j d_j=0\; .\label{zerocond}
\end{align}
After explicitly substituting in the four values for $\mathbf{y}_j$ given in \eqref{eq:P1PointsCICY6804}, we find
\begin{align}
\begin{split}
d_0&=(a_0 w_{2,0}^3+a_1 w_{2,0}^2+a_2 w_{2,0}+a_3) (c_0 w_{2,0}^3+c_1 w_{2,0}^2+c_2 w_{2,0}+c_3)-(b_0 w_{2,0}^3+b_1 w_{2,0}^2+b_2 w_{2,0}+b_3)^2\,,\\
d_1&=(a_0 w_{2,0}^3+a_1 w_{2,0}^2+a_2 w_{2,0}+a_3) (c_0 w_{2,0}^3+c_1 w_{2,0}^2+c_2 w_{2,0}+c_3)-(b_0 w_{2,0}^3+b_1 w_{2,0}^2+b_2 w_{2,0}+b_3)^2\,,\\
d_2&=(a_0 w_{2,0}^3-a_1 w_{2,0}^2+a_2 w_{2,0}-a_3) (c_0 w_{2,0}^3-c_1 w_{2,0}^2+c_2 w_{2,0}-c_3)-(b_0 w_{2,0}^3-b_1 w_{2,0}^2+b_2 w_{2,0}-b_3)^2\,,\\
d_3&=(a_0 w_{2,0}^3-a_1 w_{2,0}^2+a_2 w_{2,0}-a_3) (c_0 w_{2,0}^3-c_1 w_{2,0}^2+c_2 w_{2,0}-c_3)-(b_0 w_{2,0}^3-b_1 w_{2,0}^2+b_2 w_{2,0}-b_3)^2\; .
\end{split}
\label{eq:DeterminantsUpstairs}
\end{align}
Since these are polynomials in the 26 independent monomials $\{a_k c_l\}\cup\{b_k b_l\}$, we can formulate the vanishing condition~\eqref{zerocond} in terms of a linear system, $B\,{\bf k}=0$, where ${\bf k}=(k_0,k_1,k_2,k_3)^T$ and $B$ is a $26\times4$ matrix. It turns out that the rank of $B$ is two and that the linear system and, hence, Eq.~\eqref{zerocond} does indeed have a non-trivial solution. Note that this is only the case if the correct locations~\eqref{eq:P1PointsCICY6804} are inserted. For a generic choice of four points, the matrix $B$ has full rank. This confirms the expected vanishing of the superpotential upstairs.
\begin{figure}[t]
\centering
\includegraphics[height=1cm]{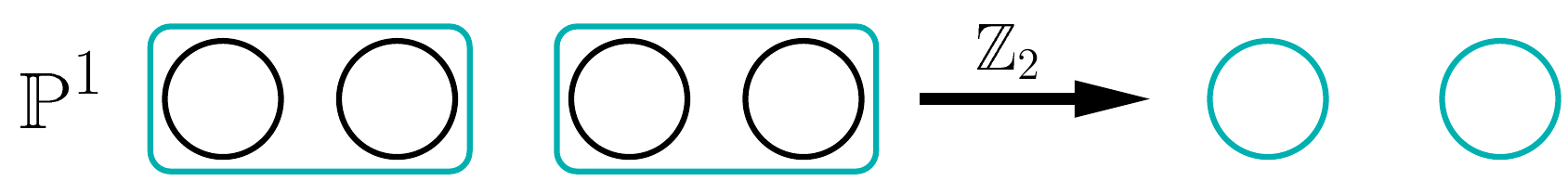}
\caption{\sf Schematic description of the action $\Gamma$ on the 4 curves in the curve class of the first $\P^1$.}
\label{fig:Example2}
\end{figure}

Let us check what happens after we mod out the freely-acting symmetry $\Gamma=\mathbbm{Z}_2$ with generator~\eqref{Z2gen}. Note that this symmetry acts by a combination of the types \ref{enum:ActionType1} to \ref{enum:ActionType3}. The schematic action of $\Gamma$ on the four curves in the curve class associated to the first $\P^1$ is given in Figure~\ref{fig:Example2}. The four curves group into two orbits of two curves each, resulting in two curves on the quotient manifold $X=\tX/\Gamma$. More specifically, it is easy to see from the generator~\eqref{Z2gen} that the four points~\eqref{eq:P1PointsCICY6804} are mapped as
\begin{equation}
 {\bf y}_0\longleftrightarrow {\bf y}_3\;,\qquad  {\bf y}_1\longleftrightarrow {\bf y}_2\; .
\end{equation} 

As mentioned before, the vector bundle $\tV$ has an equivariant structure with respect to $\Gamma$ and, hence, descends to a bundle $V$ on $X$. Upon modding out $\Gamma$ we have to restrict the extension space \eqref{eq:BasisH1} accordingly, which demands that
\begin{align}
\begin{split}
 f_1(\mathbf{y})&=a_1 y_0^2 y_1 + a_3 y_1^3\,,\\
 f_2(\mathbf{y})&=b_0 y_0^3 + b_2 y_0 y_1^2\,.\\
 f_3(\mathbf{y})&=c_1 y_0^2 y_1 + c_3 y_1^3\,.
 \end{split}
\end{align} 
Using these special expressions, the Pfaffians become
\begin{align}
d_0=d_1=d_2=d_3=(a_3 + a_1 w_{2,0}^2) (c_3 + c_1 w_{2,0}^2) - (b_2 w_{2,0} + b_0 w_{2,0}^3)^2\,.
\end{align}
The Pfaffians in each $\mathbb{Z}_2$ orbit should be proportional and, hence, the equalities $d_0=d_3$ and $d_1=d_2$ are not surprising. However, the fact that all four Pfaffians are equal allows for, and strongly hints at, a cancellation of the two downstairs contributions.

Finally, we have found the same behavior for a number of other quotients of favorable CICY manifolds with SU(3) extension bundles.

%%%%%%%%%%%%%%%%%%%%%%%%%%%%%%%%%%%%%%%%%%%%%%%%%%%%%%%%%%%%%%%%%%%%%%%%%%%
\section{Conclusions and Outlook}\label{sec:Conclusions}
We have described and illustrated a method to construct all holomorphic, isolated, genus zero curves in homology classes associated to ambient space $\mathbb{P}^1$ factors for complete intersection CY manifolds (CICY manifolds). The relevant genus zero curves can be found explicitly by a complete intersection in the ambient space. Using the traditional way of calculating the Gromov-Witten invariants via mirror symmetry, we have checked for all 7890 CICY manifolds in the standard list of Ref.~\cite{Candelas:1987kf} that our method, where applicable, indeed reproduces all genus zero curves within a given curve class. Based on this result, we conjecture that our method works for all CICY manifolds. The advantage of this approach is that it provides the curves explicitly and thereby facilitates the calculation of instanton superpotentials, both for CICY manifolds and for their quotients by discrete symmetries.

As a first application, we have identified one CICY manifold whose quotient by a freely-acting $\mathbb{Z}_4$ symmetry has only a single curve in a certain homology class. Since the superpotential contribution of this curve class is non-zero, and since there are no other curves in the same curve class that could give rise to a cancellation along the lines of Beasley-Witten, this shows that the instanton superpotential contribution from this homology class must be non-vanishing. For this example, the free $\mathbb{Z}_4$ action is non-toric, the quotient CY is not a CICY manifold, and neither the upstairs nor the downstairs CY are favorable. All three of these conditions were assumed in the original proof of the vanishing theorem of Beasley and Witten. 

In order to identify which of these conditions is crucial to avoid the vanishing result by Beasley and Witten, we studied a second example. In this case, the underlying CICY is favorable, but the freely-acting $\mathbb{Z}_2$ symmetry under consideration is non-toric and the quotient CY is not a CICY manifold. It turns out that the quotient CY has two curves in a certain homology class whose contributions to the superpotential cancel each other. We have confirmed this behavior for a number of similar examples. This suggests that the crucial property to avoid the Beasley-Witten vanishing result is the non-favorability of the CY manifold. Further, it hints at an extension of the vanishing result to quotients of favorable CICY manifolds (and complete intersections in toric spaces).

For future work, there are several avenues that are worthwhile exploring. First, it would be very interesting to extend the proof of Beasley-Witten to quotient manifolds of the type represented by our second example. It is expected that, for this class of CY manifolds, techniques similar to those used in Ref.~\cite{Beasley:2003fx} will eventually also lead to a vanishing theorem from a contour integral over a compact moduli space.

Worldsheet instanton contributions to the superpotential may be crucial for moduli stabilization and, if this is the case, it is important to avoid a cancellation \`a la Beasley and Witten. Given our results, this motivates searching for CICYs that are not favorable in any realization and that allow for a freely-acting symmetry.

It would be interesting to find an algebraic reason for the Beasley-Witten vanishing theorem. Using the methods introduced in this paper allows one to write the curves as solutions to a system of polynomial equations and the Pfaffians as another set of polynomials. In algebraic terms, the vanishing theorem means that the vanishing locus of the Pfaffians (which can be considered as a variety parametrized by complex structure and bundle moduli) always intersects the zero-dimensional variety associated to the locations of the genus zero curves. 

Finally, it is important to study the consequences of the Beasley-Witten result in the context of other approaches to string compactification. Using, for example, heterotic/F-Theory duality it would be interesting to study its implications for instanton superpotentials in F-theory or type IIB string theory.

%%%%%%%%%%%%%%%%%%%%%%%%%%%%%%%%%%%%%%%%%%%%%%%%%%%%%%%%%%%%%%%%%%%%%%%%%%%%
\section*{Acknowledgments}
We would like to thank Lara Anderson, James Gray, Thomas Grimm, Ling Lin, Irene Valenzuela and Stefan Vandoren for helpful discussions.
The work of E.I.B.\ was supported by the ARC Future Fellowship FT120100466. E.I.B.\ would like to thank physics department at Oxford University where some of this work was done for hospitality. A.L.\ would like to acknowledge support by the STFC grant ST/L000474/1. The work of A.L.\ and F.R.\ is supported by the EPSRC network grant EP/N007158/1. A.L.\ and F.R.\ thank the University of Pennsylvania for hospitality while finalizing the project. B.A.O. is supported in part by the US Department of Energy under DOE contract No. DE-SC0007901 and acknowledges support from the EPSRC network grant EP/N007158/1 for his stay at Oxford University during which this project was conceived and research work begun.
%%%%%%%%%%%%%%%%%%%%%%%%%%%%%%%%%%%%%%%%%%%%%%%%%%%%%%%%%%%%%%%%%%%%%%

\appendix

\section{Proof that the \texorpdfstring{$\boldsymbol{\P^1}$}{P1} curves are isolated}
\label{sec:IsolatedCurves}
We will prove that the $\P^1$ curves we obtain with our prescription are indeed isolated by generalizing the method of Ref.~\cite{Buchbinder:2016rmw}. To do this, we need to show that the normal bundle $NC$ of the genus zero curve $C$ within the CICY three-fold $\tX$ is
\begin{equation}
 NC={\cal O}_{\mathbb{P}^1}(-1)^{\oplus 2}\; .
\end{equation} 
Given the inclusions $C\subset \tX\subset{\cal A}$, the normal bundle $NC$ can be calculated from the two exact sequences
\begin{align}
 0\xrightarrow{~} T\tX\xrightarrow{h^{(2)}}T\mathcal{A}|_{\tX}\xrightarrow{h^{(1)}}N\tX\xrightarrow{~}0\,,\label{eq:SESCY}\\
 0\xrightarrow{~} TC\xrightarrow{\phantom{h^{(2)}}}T\tX|_C\xrightarrow{\phantom{h^{(1)}}}NC\xrightarrow{~}0\; ,\label{eq:SESCurve}
\end{align}
once the other bundles in those sequences are known. We know that $TC=\cO_C(2)$ for a $\P^1$ curve and we can obtain $T\tX$ from the first short exact sequence by studying $T\mathcal{A}|_{\tX}$, $N\tX$, and the maps $h^{(1)}$ and $h^{(2)}$. Given that, by convention, the curve $C$ is associated to the first projective factor of the ambient we have $T\mathcal{A}|_X=\cO_C(2)\oplus\bigoplus_{i=2}^{m}T\P^{n_i}$ (where we have used again that $TC=\cO_C(2)$). Furthermore, the map $h^{(1)}$ is given in terms of derivatives of the defining equation with respect to the affine coordinates of the $\P^{n_i}$. Since there are $(K+3)$ such coordinates and $K$ equations that define the CICY, $h^{(1)}$ can be represented by a $K\times(K+3)$ matrix. Similarly, since $T\tX$ is three-dimensional for a CY threefold, the map $h^{(2)}$ is given in terms of a $(K+3)\times3$ matrix. Exactness of the sequence implies $h^{(1)}\circ h^{(2)}=0$.

Since we are ultimately interested in the bundle $T\tX|_C$, we can restrict the first short exact sequence \eqref{eq:SESCY} to $C$. This leads to
\begin{align}
 0\xrightarrow{~} T\tX|_C\xrightarrow{h^{(2)}|_C}\cO_C(2)\oplus\bigoplus_{i=2}^{K+3}\cO_C\xrightarrow{h^{(1)}|_C}N\tX|_C\xrightarrow{~}0\,.
\end{align}
The last term $N\tX|_C$ can be simply read off from the configuration matrices \eqref{conf2}:
\begin{align}
\text{type $1$:~~~~} N\tX|_C=\cO_C(1)\oplus\cO_C(1)\oplus\bigoplus_{i=3}^{K}\cO_C\,,\qquad
\text{type $2$:~~~~} N\tX|_C=\cO_C(2)\oplus\bigoplus_{i=2}^{K}\cO_C\,.
\end{align}

For the maps $h^{(1)}|_C$ we get for the two types
\begin{align}
\text{type $1$:~~} h^{(1)}|_C=\begin{pmatrix}
0\!&l_{1,2}&\!\ldots\!&l_{1,K+3}\\
0\!&l_{2,2}&\!\ldots\!&l_{2,K+3}\\
0\!&\kappa_{3,2}&\!\ldots\!&\kappa_{3,K+3}\\
\vdots\!&\vdots&\!\ddots\!&\vdots\\
0\!&\kappa_{K,2}&\!\ldots\!&\kappa_{K,K+3}
\end{pmatrix}\!,\qquad
\text{type $2$:~~} h^{(1)}|_C=\begin{pmatrix}
0\!&q_{1,2}&\!\ldots\!&q_{1,K+3}\\
0\!&\kappa_{2,2}&\!\ldots\!&\kappa_{2,K+3}\\
\vdots\!&\vdots&\!\ddots\!&\vdots\\
0\!&\kappa_{K,2}&\!\ldots\!&\kappa_{K,K+3}
\end{pmatrix}\,.
\end{align}
Note that in both cases the first column $(h^{(1)}|_C)_{a,1}=\partial p_a/\partial \hat x|_C$ vanishes, where $\hat x$ is the affine coordinate of $[x_0:x_1]$ in a given patch. For those $p_a$ that do not depend on the coordinates $[x_0:x_1]$ this is true trivially, while for those equations that do depend on $[x_0:x_1]$ this is true since they vanish by construction when restricted to $C$. By the same token, the type 1 matrix contains linear\footnote{After restricting to $C$ by substituting for $\mathcal{y}$ the solution to the equations \eqref{case1} or \eqref{case2}.} polynomials $l_{1,\alpha}(x_0,x_1,\mathbf{y})$ and $l_{2,\alpha}(x_0,x_1,\mathbf{y})$, $\alpha=2,\ldots,K+3$ in the first two rows and constant polynomials $\kappa_{a,\alpha}(\mathbf{y})$ in the remaining $K-2$ rows. Similarly, for type 2, we get quadratic polynomials $q_{1,\alpha}(x_0,x_1,\mathbf{y})$ in the first row and constant polynomials $\kappa_{a,\alpha}(\mathbf{y})$ in the remaining $K-1$ rows. 

In order to find $T\tX|_C$ we next study the $(K+3)\times 3$ matrix $h^{(2)}|_C=(h)_{\alpha,d}$, $\alpha=1,\ldots,K+3$, $d=1,2,3$. From $h^{(1)}\circ h^{(2)}=0$ we get $3K$ equations. Note that since the first column of $h^{(1)}|_C$ is zero the first row $h^{(2)}_{1,d}$ is not fixed and we will deal with these entries separately. We first focus on the other entries $h^{(2)}_{\alpha,d}$, $\alpha=2,\ldots,K+3$. In general these are polynomials in $x_0$ and $x_1$. In order to determine their rank we study how many coefficients we need in order to satisfy $h^{(1)}|_C\circ h^{(2)}|_C=0$ for non-trivial $h^{(2)}|_C$. Since the discussion of the two types proceeds in a slightly different way we include separate discussions.

\subsubsection*{Type 1 curves}
Let us start with type 1.  In the simplest case the polynomials $h^{(2)}_{\alpha,d}$, $\alpha=2,\ldots,K+3$, $d=1,2,3$ are just constants. There are $3(K+2)=3K+6$ of them. In the first $3\cdot 2$ equations we need to choose the $h^{(2)}_{\alpha,d}$ such that the corresponding sums of the linear polynomials vanish identically, i.e.\ the coefficient in front of the $x_0$ and $x_1$ terms have to be zero. For the remaining $3K-6$ equations we need to ensure that the corresponding $\kappa_{a,\alpha}$ sum to zero. Together we thus get $3\cdot2\cdot2+3K-6=3K+6$ conditions on the $3K+6$ coefficients, which has a unique solution. However, since the system of equations is homogeneous this unique solution means that $h^{(2)}|_C$ is trivial, which is impossible since $T\mathcal{A}|_{\tX}$ is non-trivial. 

Hence we need the $h^{(2)}_{\alpha,d}$ to be at least linear polynomials. In that case there are $2\cdot3(K+2)=6K+12$ coefficients. For the first 6 equations to vanish identically we need the coefficients in front of $x_0^2$, $x_0x_1$ and $x_1^2$ to vanish. For the remaining $3K-6$ equations we need the coefficients in front of $x_0$ and $x_1$ to vanish. Together this gives $3\cdot2\cdot3+2(3K-6)=6K+6$ conditions on the $6K+12$ coefficients, which has non-trivial solutions.

\subsubsection*{Type 2 curves}
In order to study the curves of type 2 we proceed in a similar fashion. Again the simplest case would be to choose constant $h^{(2)}_{\alpha,d}$, $\alpha=2,\ldots,K+3$, $d=1,2,3$. There are again $3K+6$ of them. In the first 3 equations we need to choose the $h^{(2)}_{1,d}$ such that the quadratic polynomials vanish identically, which amounts to 3 constraints for each $d$ from the coefficients of $x_0^2$, $x_0x_1$ and $x_1^2$. In the remaining $3K-3$ equations we need to arrange for the $\kappa$'s to sum to zero, such that we get a total of $3\cdot3+3K-3=3K+6$ conditions on the $3K+6$ coefficients. Thus there will again only be the trivial solution in this case, which is ruled out.

Next, we try linear polynomials for the $h^{(2)}_{\alpha,d}$, $\alpha=2,\ldots,K+3$. There are $6K+12$ coefficients. Imposing the first three equations to vanish gives rise to 4 constraints per equation from the coefficients of the $x_0^3$, $x_0^2x_1$, $x_0x_1^2$, $x_1^3$ terms. In the other $3K-3$ equations we need to arrange the coefficients of the $x_0$ and $x_1$ term to cancel, leading to a total of $4\cdot3+2(3K-3)=6K+6$ constraints on the $6K+12$ coefficients, which allows for a non-trivial solution. Note that the counting in the end is the same as for type 1, albeit for different reasons.

Having established that all $h^{(2)}_{\alpha,d}$, $\alpha=2,\ldots,K+3$ are linear for both types let us now come back to the the polynomials $h^{(2)}_{1,d}$ in the first row. In order to fix them we look at the pre-image and the image of $h^{(2)}|_C$. Since $T\tX$ is three-dimensional and since every bundle on $\P^1$ can be written as a sum of line bundles we can write $T\tX|_C=\cO_C(m_1)\oplus\cO_C(m_2)\oplus\cO_C(m_3)$. Furthermore, from inspecting the first Chern class of the short exact sequence \eqref{eq:SESCY} we find that $m_1+m_2+m_3=0$. Since the action of $h^{(2)}_{\alpha,d}$ increases the $m_i$ by one, and since we only have one non-trivial element, $\cO_C(2)$, in the image, the only possibility is $m_1=2$, $m_2=m_3=-1$. In this case $h^{(2)}_{1,1}=1$, $h^{(2)}_{1,d}=h^{(2)}_{\alpha,1}=0$ for $d=2,3$, $\alpha=2,\ldots,K+3$. Thus $T\tX|_C=\cO_C(2)\oplus\cO_C(-1)\oplus\cO_C(-1)$.

In conclusion, the sequence \eqref{eq:SESCurve} becomes
\begin{align}
   0\xrightarrow{~} \cO_C(2)\xrightarrow{\phantom{h_2}}\cO_C(2)\oplus\cO_C(-1)\oplus\cO_C(-1)\xrightarrow{\phantom{h_1}}NC\xrightarrow{~}0\,.
\end{align}
The only possibility is that the sequence splits and that $NC=\cO_C(-1)^{\oplus 2}$. Hence, the curve $C$ is isolated.

%\bibliographystyle{bibstyle}
%\bibliography{references}

\providecommand{\href}[2]{#2}

\end{document}